\documentclass[12pt]{article}
\usepackage{pic02}
\usepackage{hyperref}
\usepackage{url}
\usepackage{graphicx}

\def\bbar{\bar{b}{}}
\def\sbar{\bar{s}{}}
\def\qbar{\bar{q}{}}
\def\nubar{\bar{\nu}{}}
\def\qqbar{q\qbar}
\def\Xs{X_s}
\def\Bbar{\overline{B}{}}
\def\Kbar{\overline{K}{}}
\def\Jpsi{J/\psi}

\def\btosgamma{b\to s\gamma}
\def\bbartosbargamma{\bbar\to\sbar\gamma}
\def\btodgamma{b\to d\gamma}
\def\btosll{b\to s\ell^+\ell^-}
\def\btoclnu{b\to c\ell^-\nubar}
\def\btoulnu{b\to u\ell^-\nubar}
\def\Btorhogamma{B\to \rho\gamma}
\def\Btoomegagamma{B\to \omega\gamma}
\def\BtoXsgamma{B\to X_s\gamma}
\def\BtoXsll{B\to X_s\ell^+\ell^-}

\def\BtoKll{B\to K\ell^+\ell^-}
\def\BtoKstarll{B\to K^*\ell^+\ell^-}
\def\BtoKstaree{B\to K^*e^+e^-}
\def\BtoKstarmumu{B\to K^*\mu^+\mu^-}
\def\BtoKorKstarll{B\to K^{(*)}\ell^+\ell^-}
\def\BtoKstargamma{B\to K^*\gamma}
\def\BbartoKstarbargamma{\Bbar\to \Kbar^*\gamma}

\def\BtoKstarzerogamma{B^0\to K^{*0}\gamma}
\def\BtoKstarplusgamma{B^+\to K^{*+}\gamma}
\def\Btorhozerogamma{B^0\to \rho^0\gamma}
\def\Btorhoplusgamma{B^+\to \rho^+\gamma}

\def\fbinv{\mbox{~fb}^{-1}}
\def\GeV{\mbox{~GeV}}

\def\GeVcc{\mbox{~GeV}/c^2}

\def\Acp{A_{cp}}
\def\Mbc{M_{bc}}
\def\MES{M_{ES}}
\def\DeltaE{\Delta{E}}

\def\CsevenEff{C_7^{\rm eff}}
\def\CnineEff{C_9^{\rm eff}}
\def\CtenEff{C_{10}^{\rm eff}}

\def\CnineNP{C_{9}^{\rm NP}}
\def\CtenNP{C_{10}^{\rm NP}}

\def\Mll{M_{\ell^+\ell^-}}

\def\epem{e^+e^-}
\def\mpmm{\mu^+\mu^-}
\def\MXs{M_{\Xs}}

\def\Vcb{V_{cb}}
\def\Vub{V_{ub}}
\def\Vtd{V_{td}}
\def\Vts{V_{ts}}

\def\Br{{\cal B}}

\begin{document}

\title{\bf PROBING BEYOND STANDARD MODEL PHYSICS WITH ELELCTROWEAK PENGUIN
       \boldmath $B$ DECAYS}
\author{Mikihiko Nakao \\
{\em KEK, High Energy Accelerator Research Organization,}\\
{\em Tsukuba, Ibaraki 305-0801, JAPAN}}

\maketitle

% photograph of author
%  This is where we will insert a photograph. To see what it would look like,
%  uncomment the following lines.
%
%%%%%%%%%%%%%%%%%%%%%%%%%%%%%%%%%%%%%%%%%%%%%%%%%%%%%%%%%%%%%%%%%%%%%%
%%%%%%%%%%%%%%%%%%%%%%%%%%%% INCLUDE PHOTOGRAPH FOR PROCEEDING VERSION
%%%%%%%%%%%%%%%%%%%%%%%%%%%%%%%%%%%%%%%%%%%%%%%%%%%%%%%%%%%%%%%%%%%%%%
%\begin{figure}[h]
%\begin{center}
%\includegraphics[height=4.5cm]{einstein.eps}
%\end{center}
%\end{figure}
%
% insert a fixed vertical spacing instead for the ArXiv preprint
%
\vspace{4.5cm}

%%%%%%%%%%%%%%%%%%%%%%%%%%%%%%%%%%%%%%%%%%%%%%%%%%%%%%%%%%%%%%%%%%%%%%

\baselineskip=14.5pt
\abstract{
Latest experimental progress is reviewed on the searches for physics
beyond the Standard Model using the radiative and electroweak penguin
decays of $B$ mesons.  This review covers inclusive and exclusive
measurements of the $\btosgamma$, $\btodgamma$ and $\btosll$ processes,
including the first observation of $\BtoKll$ and the first attempt to
measure the inclusive $\BtoXsll$ branching fraction.
}
\newpage
\baselineskip=17pt

%%%%%%%%%%%%%%%%%%%%%%%%%%%%%%%%%%%%%%%%%%%%%%%%%%%%%%%%%%%%%%%%%%%%%%
\section{Introduction} %%%%%%%%%%%%%%%%%%%%%%%%%%%%%%%%%%%%%%%%%%%%%%%

Since the first measurement of the inclusive $\BtoXsgamma$ decay rate by
CLEO in 1995 \cite{bib:cleo-first-btosgamma}, rare $B$ decays involving
the penguin diagram have been a unique probe to search for new physics.
The inclusive decay $\BtoXsgamma$ corresponds to the quark level process
$\btosgamma$, which is to date accurately calculated up to the
next-to-leading order (NLO) QCD corrections.  In the Standard Model
(SM), the lowest order diagram for $\btosgamma$ is a loop (radiative
penguin) diagram of top quark and $W$ boson.  In principle, new
particles such as charged Higgs or SUSY partners can form the same loop
diagram and may modify the SM amplitude.  A comparison between the
measured rate and the SM prediction has provided a stringent constraint
on such new particles.  As inclusive measurements have already been
extensively performed, an exclusive measurement of $\BtoKstargamma$ does
not give a further constraint to new physics, because the model
dependent form factor uncertainties in the SM predictions of the
exclusive channels are too large.

Similar processes, $\btodgamma$ and $\btosll$, are also useful probes
for new physics searches.  Expected rates are two order of magnitude
smaller than for $\btosgamma$, as $\btodgamma$ is suppressed by
$|\Vtd/\Vts|^2$ and $\btosll$ is suppressed by an additional factor of
$\alpha_{\rm em}$.  At the lowest order, $\btosll$ process is described
by an electroweak ($Z$) penguin diagram and a $W$-box diagram in
addition to the radiative penguin.  One can therefore expect some
additional modifications to $\btosll$ that are not visible in
$\btosgamma$, if there exist new particles that have large couplings
with weak bosons.  For these very rare decays, measurements of exclusive
modes such as $\Btorhogamma$ and $\BtoKorKstarll$ also provide useful
information even with the large uncertainties in the SM predictions,
until the inclusive branching fraction for the corresponding process is
known.

The SM amplitudes are calculated using an operator product expansion
technique.  The coefficients of the operators are called Wilson
coefficients, $C_i$, that are theoretically calculated.  At lowest
order, $\btosgamma$ is described by the size of the coefficient $C_7$.
For $\btosll$, the coefficients $C_7$, $C_9$ and $C_{10}$ contribute.
Higher order QCD corrections introduce other operators; however, one can
absorb those contributions by modifying the lowest order coefficients
into effective coefficients $\CsevenEff$, $\CnineEff$ and $\CtenEff$.
The measured $\btosgamma$ rate provides a stringent limit on
$|\CsevenEff|$, and then $\btosll$ results can be used to extract
$\CnineEff$ and $\CtenEff$, together with the sign of $\CsevenEff$.
In general, new physics can be modeled by introducing additional non-SM
components $C_i^{\rm NP}$ to these Wilson coefficients that can be
searched for by comparing the measured $C_i$ and their SM predictions.

Other observables, such as the partial rate asymmetry ($\Acp$) between
charge conjugate modes, are also useful to constrain new physics.  For
example, the SM predicts very small asymmetry, while there are several
extensions of the SM that predict much larger $\Acp$.  SM predictions
for the $\Acp$ of the exclusive channels are also reliable.

Studies of such rare decays have been pioneered by CLEO, which has
accumulated about $13\fbinv$ data.  Now, two B-factory experiments,
Belle and BaBar, have already superseded CLEO in the size of the
collected data.  All three detectors have similar experimental
capabilities and comparable sensitivities for rare decays at a given
integrated luminosity.

%%%%%%%%%%%%%%%%%%%%%%%%%%%%%%%%%%%%%%%%%%%%%%%%%%%%%%%%%%%%%%%%%%%%%%
\section{Radiative decays $\btosgamma$} %%%%%%%%%%%%%%%%%%%%%%%%%%%%%%

\begin{figure}[htb]
 \begin{center}
 \includegraphics[width=0.4\textwidth]{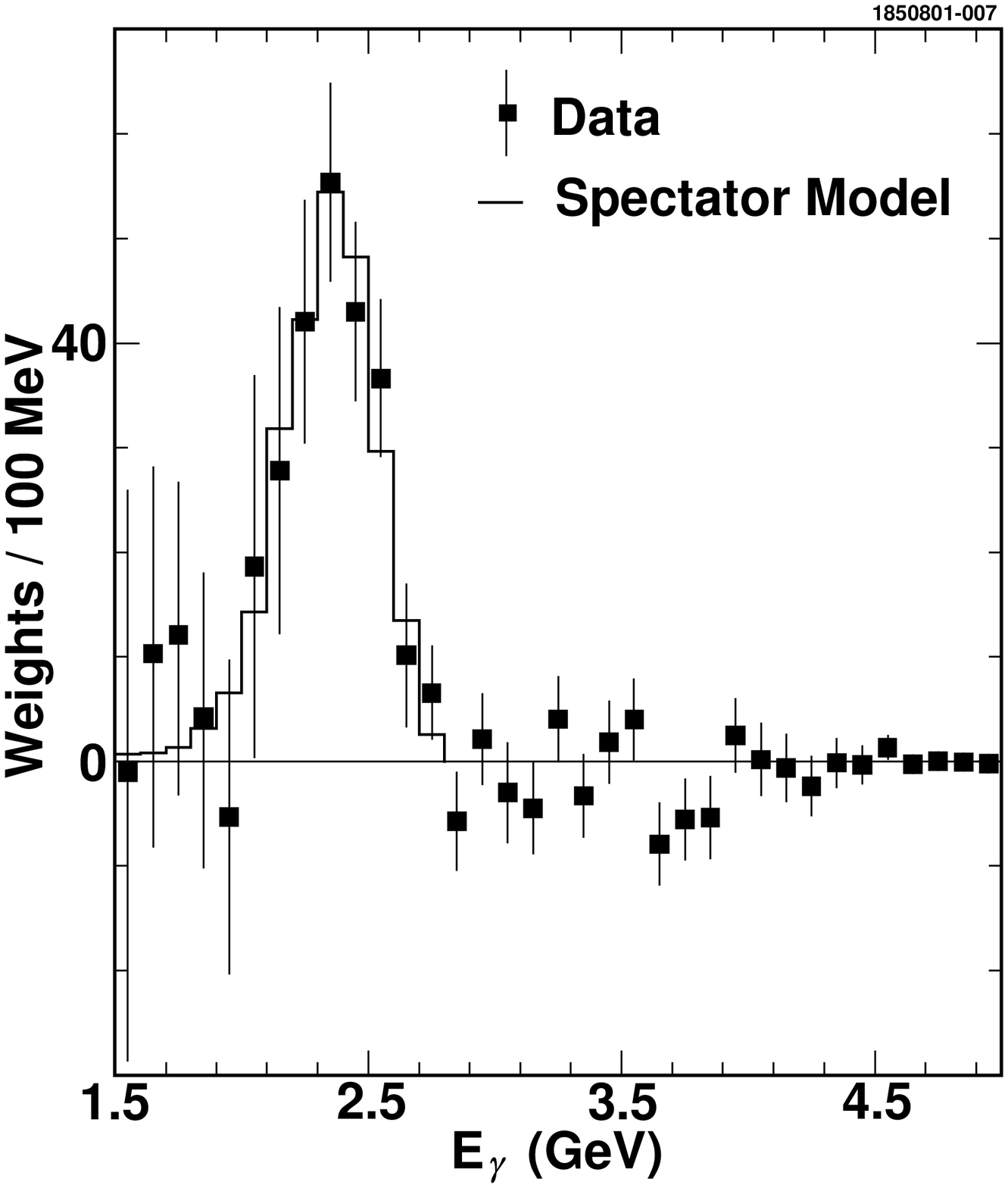}
 \caption{\it Photon energy spectrum of $\BtoXsgamma$ by CLEO.}
 \label{fig:cleo-xsgam}
 \end{center}
\end{figure}

The decay $\btosgamma$ has a clear signature of an energetic photon in
the range between $2$ to $2.7\GeV$ due to its kinematics of two-body
decay from the almost at-rest $B$ meson (Figure~\ref{fig:cleo-xsgam}).
Underlying the signal, there are large backgrounds from continuum
$\qqbar$ $(q=u,d,s,c)$ productions, in which photons originate from
initial state radiation or energetic $\pi^0$, $\eta$ and other light
mesons.  In principle, this continuum background can be subtracted by
using an off-resonance data sample taken below the $\Upsilon(4S)$
resonance.  In addition, there are $B$ decay backgrounds in the photon
energy range below $2.2\GeV$.  For the subtraction of the $B$ decay
background, one has to rely on Monte Carlo (MC).  The dominant part is
from $B\to\pi^0 X$, for which MC is tuned by using the measured $\pi^0$
spectrum from $B$ decays.  Other $B$ decay backgrounds are considerably
smaller and reasonably modeled with MC.  The recoil system $X_s$
provides another useful background discrimination.  By summing up the
combinations of one kaon and 1 to 4 pions, one can perform a
pseudo-reconstruction of the kinematic variables of $B$ decays such as
the beam-energy constrained (substituted) mass $\Mbc$ ($\MES$) and the
energy difference $\DeltaE$.  These variables are explicitly used in the
Belle analysis or included in the background suppression and candidate
selection by CLEO.

The measured $\BtoXsgamma$ branching fractions \cite{bib:exp-btosgamma}
are summarized in Table~\ref{tbl:btoxsgamma}, together with SM
predictions \cite{bib:theo-btosgamma}.  Until recently the theory error
has been considered to be about 10\%, but now there are arguments about
the uncertainty of the charm quark mass included in the higher order
loop calculation, which can be different from the pole mass as
originally assumed.  By taking into account this additional uncertainty,
the overall uncertainty is about 15\%.  The measurement error of the
world average is about 12\%, and the branching fraction is in good
agreement with the SM expectation.  In addition to the branching
fraction, the measured photon energy spectrum provides information about
$B$ meson decay dynamics using the framework of heavy quark effective
theory (HQET).  This information has found to be quite useful to
extrapolate the $\btoclnu$ and $\btoulnu$ measurements to the entire
phase space in a less model dependent way, and to provide reliable
values of $|\Vcb|$ and $|\Vub|$ \cite{bib:marsiske}.

\begin{table}[ht]
\caption{\it Measured and predicted branching fractions for $\BtoXsgamma$.}
\label{tbl:btoxsgamma}
\begin{center}
\begin{tabular}{ll}
\hline
           & Branching Fraction $(\times 10^{-4})$ \\
\hline
CLEO 2001  &
  $3.21 \pm 0.43{\rm(stat)} \pm 0.27{\rm(syst)} ^{+0.18}_{-0.10}{\rm(th)}$ \\
Belle 2001 &
  $3.36 \pm 0.53{\rm(stat)} \pm 0.42{\rm(syst)} ^{+0.50}_{-0.54}{\rm(th)}$ \\
ALEPH 1998 &
  $3.11 \pm 0.80{\rm(stat)} \pm 0.72{\rm(syst)}$ \\
Average of measurements   & $3.22 \pm 0.40$ \\
\hline
Chetyrkin et al. 1997,  $m_c({\rm pole})$            & $3.28\pm0.33$ \\
Buras et al. 2002, $m_c(\overline{MS}(\mu))$ & $3.57\pm0.30$ \\
\hline
\end{tabular}
\end{center}
\end{table}

The partial rate asymmetry between $\btosgamma$ and $\bbartosbargamma$
is predicted to be less than 1\% in the SM, and therefore any asymmetry
beyond this will be a clear sign of new physics.  This has been measured
by CLEO \cite{bib:cleo-acp-btosgamma}, to be
$\Acp=-0.079\pm0.108\pm0.022$, which is consistent with no asymmetry.

%%%%%%%%%%%%%%%%%%%%%%%%%%%%%%%%%%%%%%%%%%%%%%%%%%%%%%%%%%%%%%%%%%%%%%
\section{Exclusive radiative decays} %%%%%%%%%%%%%%%%%%%%%%%%%%%%%%%%%

\begin{figure}[htb]
 \begin{center}
 \includegraphics[width=0.65\textwidth]{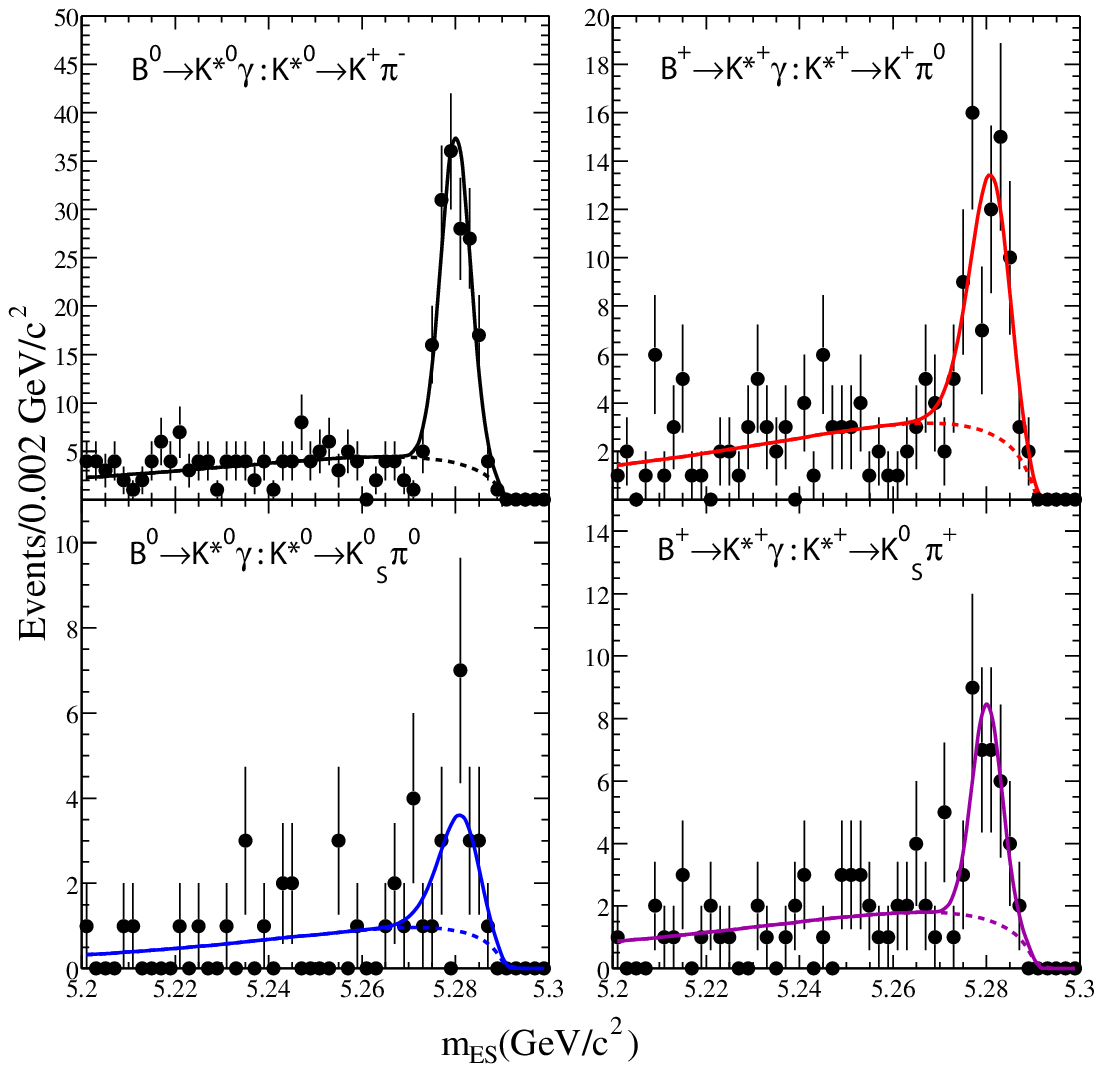}
 \caption{\it $\BtoKstargamma$ signal by BaBar.}
 \label{fig:babar-kstargam}
 \end{center}
\end{figure}

In contrast to the inclusive measurement, reconstruction of exclusive
channels such as $\BtoKstargamma$ is fairly easy.  Once the candidate is
identified with $\Mbc$ $(\MES)$ and $\DeltaE$, the continuum background
can be suppressed to a low level by using standard techniques such as
cuts on $R_2$ or $\cos\theta_{\rm thrust}$
(Figure~\ref{fig:babar-kstargam}).  Backgrounds from other $B$ decays
are even smaller.  The $\BtoKstargamma$ branching fractions reported by
CLEO, BaBar and Belle \cite{bib:exp-kstargamma} are becoming very
accurate as the data samples become large.  However, corresponding
theoretical predictions \cite{bib:theo-kstargamma} suffer from large
uncertainties, and as a result, exclusive modes are not as useful as
the inclusive measurement to constrain new physics.  The latest results
are summarized in Table~\ref{tbl:btokstargamma}.  The measured branching
fractions are very precise, and the error may not shrink rapidly as the
size of the statistical error has already reached that of the systematic
error.

\begin{table}[ht]
  \caption{\it Measured and predicted branching fractions for
               $\BtoKstargamma$.}
  \label{tbl:btokstargamma}
  \begin{center}
  \begin{tabular}{lcc}
  \hline
  & \multicolumn{2}{c}{Branching Fraction $(\times 10^{-5})$} \\
  & $\BtoKstarzerogamma$ & $\BtoKstarplusgamma$ \\
  \hline
  CLEO  2000  $(9\fbinv)$ & $4.55\pm0.70\pm0.34$ & $3.76\pm0.86\pm0.28$ \\
  BaBar 2002 $(21\fbinv)$ & $4.23\pm0.40\pm0.22$ & $3.83\pm0.62\pm0.22$ \\
  Belle preliminary $(29\fbinv)$& $4.08\pm0.34\pm0.26$ & $4.92\pm0.57\pm0.38$\\
  Average of measurements   & $4.21\pm0.25\pm0.26$ & $4.32\pm0.38\pm0.30$ \\
  \hline
  \hbox to 0pt{Ali, Parkhomenko  2002 (Large Effective Energy Theory)\hss}
        & \multicolumn{2}{c}{$7.2\pm2.7$} \\
  Bosch, Buchalla 2002 (QCD factorization) & \multicolumn{2}{c}{$7.1\pm2.5$} \\
  \hline
  \end{tabular}
  \end{center}
\end{table}

The rate difference between neutral and charged $\BtoKstargamma$ also
provides constraints on new physics.  The latest results do
not show such a difference.

The decay $\BtoKstargamma$ accounts for 13\% of total inclusive
$\btosgamma$ decays.  Exclusive decays through higher resonances
provides additional information on the $\Xs$ system
\cite{bib:belle-kxgamma}.  In the $K\pi\gamma$ final state, CLEO and
Belle find evidence for $B\to K_2^*(1430)\gamma$.  Belle searched for
other exclusive channels in the $B^+\to K^+\pi^-\pi^+\gamma$ final
state.  So far no particular resonant state is disentangled; however,
$B^+\to K^{*0}\pi^+\gamma$ and $B^+\to K^+\rho^0\gamma$ branching
fractions are measured separately.  The results are summarized in
Table~\ref{tbl:btokxgamma}.  These results are used to estimate the
decay rates into the unmeasured charge combinations of $K^*\pi\gamma$
and $K\rho\gamma$ assuming isospin.  It is found that $35\pm8\%$ of
inclusive rate is accounted for by the exclusive decays with $K\pi$ and
$K\pi\pi$ final states.  The remainder must be accounted for by the
final states with $\eta$ or $\eta'$ mesons, more than 2 pions, more than
1 kaon, or final states with baryons.

\begin{table}[ht]
  \caption{\it Branching fractions for exclusive radiative decays
  other than $\BtoKstargamma$.}
  \label{tbl:btokxgamma}
  \begin{center}
  \begin{tabular}{lc}
  \hline
  & Branching Fraction $(\times 10^{-5})$ \\
  \hline
  CLEO $B\to K_2^*(1430)\gamma$ (No $K^*(1410)$ is assumed)
                   & $1.66\pm0.56\pm0.13$\\
  Belle $B^0\to K_2^*(1430)^0\gamma$ & $1.5\pm0.6\pm0.1$ \\
  \hline
  Belle $B^+\to K^{*0}\pi^+\gamma$ & $2.0^{+0.7}_{-0.6}\pm0.2$\\
  Belle $B^+\to K^+\rho^0\gamma$   & $1.0\pm0.5^{+0.2}_{-0.1}$\\
  Belle $B^+\to K^+\pi^-\pi^+\gamma$ (N.R.) &  $<0.9$ (90\% CL)\\
  \hline
  \end{tabular}
  \end{center}
\end{table}

The partial rate asymmetry between $\BbartoKstarbargamma$ and
$\BtoKstargamma$ in the SM is expected to be as small as in the
inclusive case.  The asymmetries measured by CLEO, BaBar and Belle are
all consistent with no asymmetry, or in average, $\Acp =
(+0.9\pm4.8\pm1.8)\times10^{-2}$.  Here, the statistical error is still
dominant, and the error will be reduced to the level of 1--2\% in the
near future to provide a stringent constraint on new physics.

%%%%%%%%%%%%%%%%%%%%%%%%%%%%%%%%%%%%%%%%%%%%%%%%%%%%%%%%%%%%%%%%%%%%%%
\section{$\btodgamma$ process} %%%%%%%%%%%%%%%%%%%%%%%%%%%%%%%%%%%%%%%

Similarly to the $\btosgamma$ process, $\btodgamma$ is sensitive to new
physics.  The expected branching ratio $\Br(\btodgamma)/\Br(\btosgamma)$
is approximately proportional to $|\Vtd/\Vts|^2$.  As the size of
$|\Vtd|$ is poorly known, $\btodgamma$ modes are useful for the
determination of $|\Vtd/\Vts|$ until it is determined by the Tevatron
experiments using the anticipated measurement of $\Delta m_s/\Delta m_d$.

No inclusive measurement of $\btodgamma$ has been attempted so far.  An
exclusive measurement of $\Btorhogamma$ or $\Btoomegagamma$ is in
principle a copy of the $\BtoKstargamma$ measurement.  The main issues
are the more severe continuum background due to much lower branching
fractions and the background from mis-identified $\BtoKstargamma$ in
$\Btorhogamma$.  Recently the BaBar group reported a significantly
improved upper limit on $\Btorhogamma$ \cite{bib:babar-rhogamma}, by
improving the pion selection algorithm.  The reported results are,
$\Br(\Btorhozerogamma)<1.5\times10^{-6}$ and
$\Br(\Btorhoplusgamma)<2.8\times10^{-6}$ at 90\% confidence level.
These limits are still slightly above the expected SM branching
fractions, and do not provide an additional constraint on $|\Vtd|$.

%%%%%%%%%%%%%%%%%%%%%%%%%%%%%%%%%%%%%%%%%%%%%%%%%%%%%%%%%%%%%%%%%%%%%%
\section{Observation of $\BtoKorKstarll$} %%%%%%%%%%%%%%%%%%%%%%%%%%%%

One of the recent highlights in $B$ physics is the observation of the
$\btosll$ process, which has only become possible in the $B$-factory
era.  CLEO, CDF and other experiments have previously searched for the
decay $\BtoKorKstarll$ without success.

\begin{figure}[htb]
 \begin{center}
 \hbox{
   \includegraphics[width=0.47\textwidth]{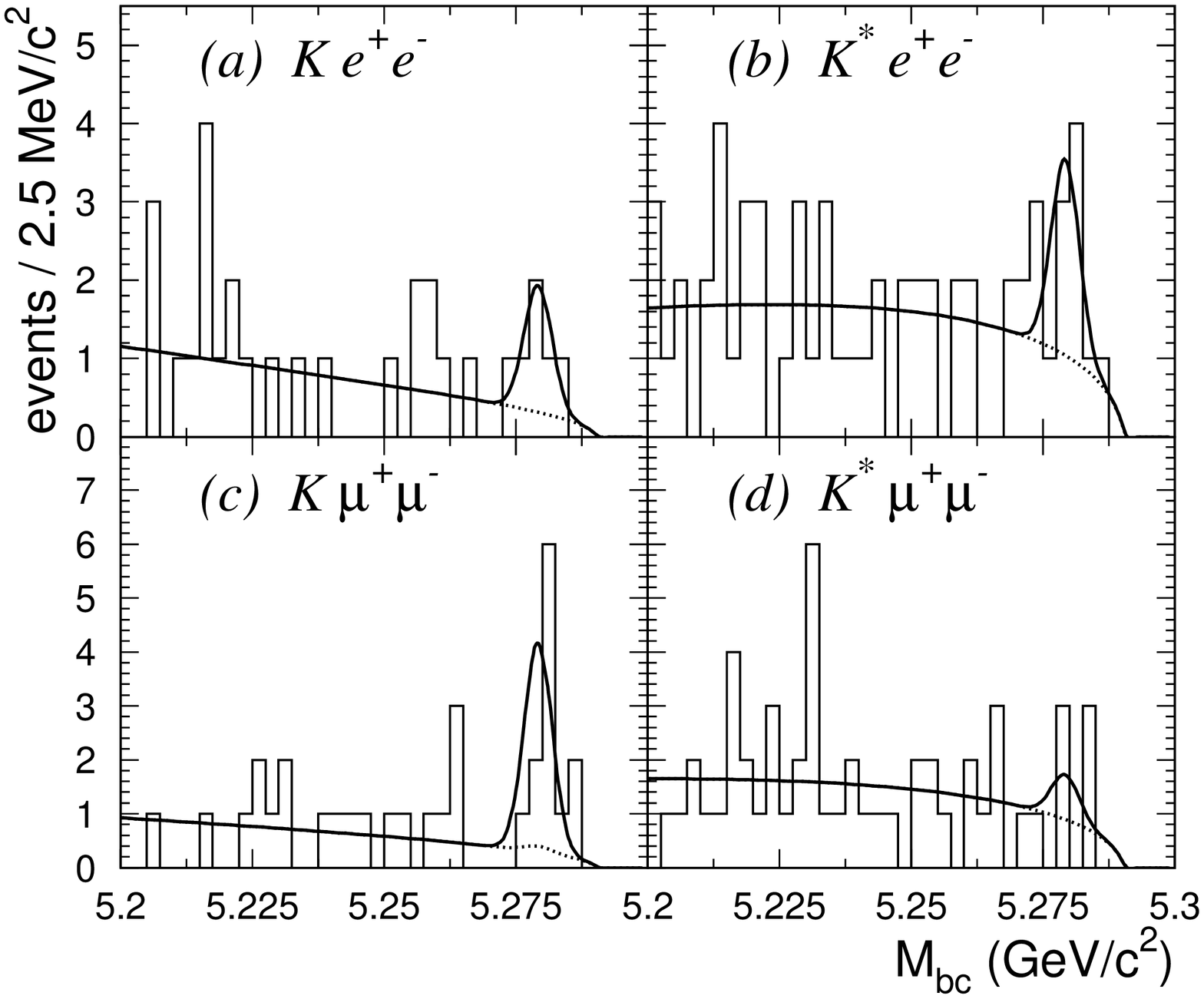}
   \hspace{0.05\textwidth}
   \raisebox{-24pt}{\includegraphics[width=0.47\textwidth]{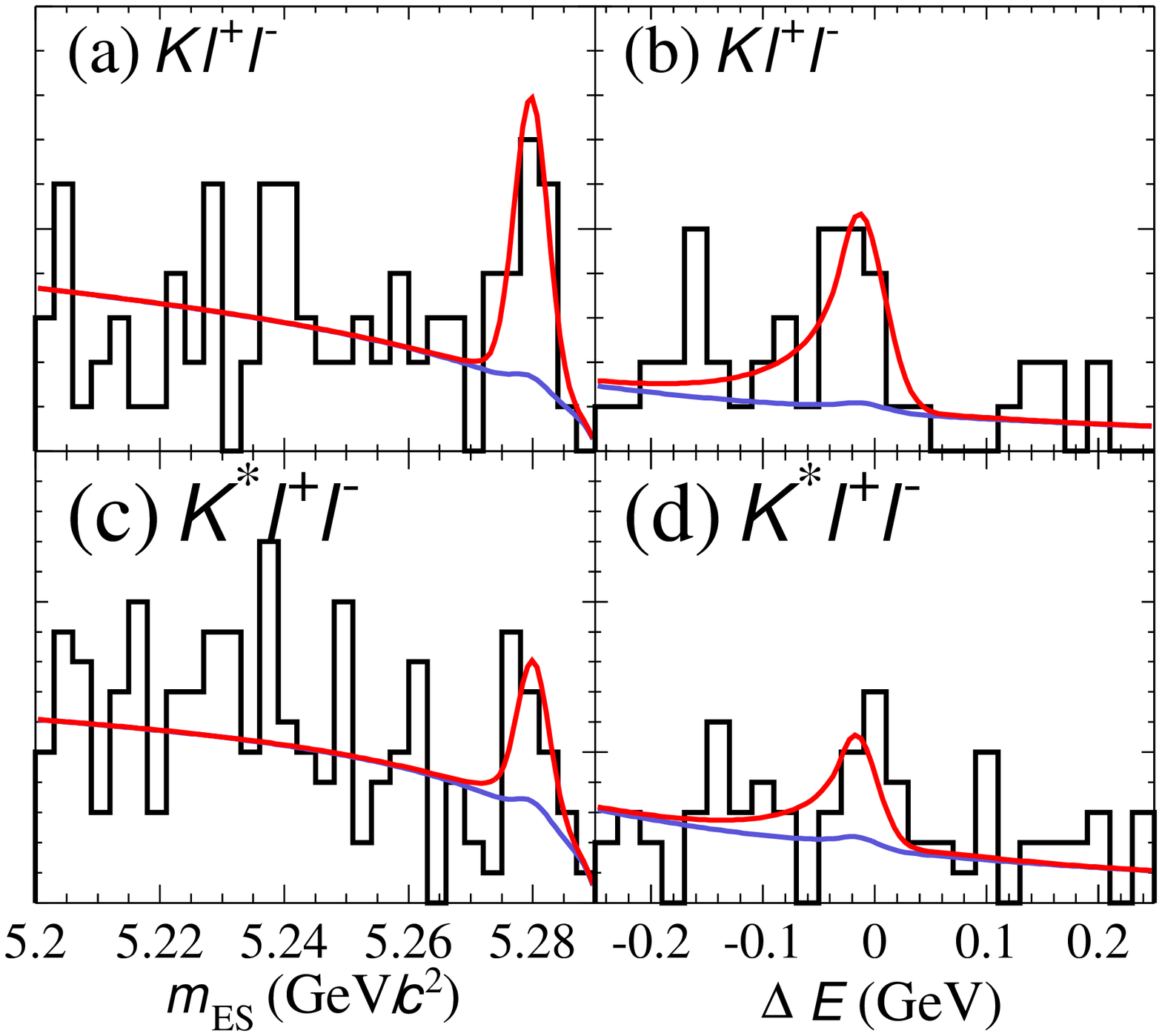}}}
 \caption{\it $\BtoKll$ signal and some hints for $\BtoKstarll$ by Belle
          (left) and BaBar (right).}
 \label{fig:korkstarll}
 \end{center}
\end{figure}

The first observation of the decay $\BtoKll$ has been made by Belle,
using $29\fbinv$ of data \cite{bib:belle-kll}.  A lepton pair with an
additional kaon is a very clear signal; however, there are a number of
large sources of background.  The largest background is due to the
oppositely charged two leptons from semileptonic decays of both of the
$B$ meson pair or a $b\to c\ell\nu$ decay with a cascade $c\to s\ell\nu$
decay.  The continuum background is suppressed by using shape variables.
Electron pairs with small invariant masses are removed to reject
$\pi^0\to\epem\gamma$ and $\gamma^*\to\epem$ conversions.  The
charmonium decays $B\to\Jpsi K$ and $B\to\psi'K$ have the same final
states and interfere with the $\BtoKll$ signal.  For this analysis, the
$\Mll$ regions around the $\Jpsi$ and $\psi'$ masses are removed.  The
removed area is much wider in the lower mass side, and especially for
electrons, to account for energy loss due to the photon radiation from
the electrons.  When two pions are mis-identified as leptons, the
copious $B\to K\pi\pi$ events from $B\to D\pi$ cannot be distinguished.
The double mis-identification probability is very small.  As a result
the background from mis-identification is only 0.3 event for $K\mpmm$
and much less for $K\epem$.  These backgrounds are subtracted from the
signal yield.  Belle observed 13.6 $\BtoKll$ signal events with a
$5.3\sigma$ significance (Figure~\ref{fig:korkstarll}-left) and obtained
a branching fraction of $(7.5^{+2.5}_{-2.1}\pm0.9)\times10^{-7}$.

A similar analysis has been performed by the BaBar group.  Initially
BaBar did not observe the $\BtoKll$ signal using $20\fbinv$ data, but
the $\BtoKll$ signal is now observed with an updated dataset of
$56\fbinv$ \cite{bib:babar-kll} (Figure~\ref{fig:korkstarll}-right).
Both results are consistent with the Belle results.  Belle and BaBar
have also searched for $\BtoKstarll$, but so far no significant signal
is observed.  The results are summarized in
Table~\ref{tbl:btokorkstarll}.

\begin{table}[ht]
  \caption{\it Latest results for $\BtoKorKstarll$.}
  \label{tbl:btokorkstarll}
  \begin{center}
  \begin{tabular}{lccc}
  \hline
   & significance & branching fraction & upper limit \\
   &              & ($\times10^{-7}$)  & ($\times10^{-7}$, 90\% C.L.) \\
  \hline
  Belle $\BtoKll$ & $5.3\sigma$ & $7.5^{+2.5}_{-2.1}\pm0.9$ \\
  BaBar $\BtoKll$ & $5.0\sigma$ & $8.4^{+3.0}_{-2.4}{}^{+1.0}_{-1.8}$\\
  \hline
  BaBar $\BtoKstarll$   & $3.5\sigma$ & $18.9^{+8.4}_{-7.2}\pm3.1$
                                    & $<35$ \\
  Belle $\BtoKstaree$   & $2.5\sigma$ & $20.8^{+12.3}_{-10.0}{}^{+3.5}_{-3.7}$
                                    & $<56$\\
  Belle $\BtoKstarmumu$ &  ---        & --- & $<31$ \\
  \hline
  \end{tabular}
  \end{center}
\end{table}

The results for $\BtoKorKstarll$ have been used to constrain non-SM
contributions $\CnineNP$ and $\CtenNP$ to the Wilson coefficients.  The
upper limit results have been excluding the outer part of a circular
area on the $\CnineNP$--$\CtenNP$ plane.  The non-zero branching
fraction results are now used to exclude the inner part of the
$\CnineNP$--$\CtenNP$ plane for the first time \cite{bib:ali-2002}.

%%%%%%%%%%%%%%%%%%%%%%%%%%%%%%%%%%%%%%%%%%%%%%%%%%%%%%%%%%%%%%%%%%%%%%
\section{Inclusive $\BtoXsll$} %%%%%%%%%%%%%%%%%%%%%%%%%%%%%%%%%%%%%%

In order to reduce the theoretical error, it is desirable to have an
inclusive measurement of $\btosll$.  Belle has attempted a
pseudo-reconstruction of $\BtoXsll$, where $\Xs$ is reconstructed as one
kaon and 0 to 4 pions, of which up to one $\pi^0$ is allowed.
The background reduction conditions are tighter than in the exclusive
analysis.  In addition, the mass of $\Xs$ is required to be less than
$2.1\GeVcc$ to reduce the large combinatorial background.  The
mis-identification background is more severe, since $B\to\Xs\pi^+\pi^-$
includes many decay channels of $B\to Dn(\pi)$ $(n\ge1)$ with large
branching fractions.  The expected background yield is $2.4\pm0.4$
events, which is subtracted from the signal yield.

\begin{figure}[htb]
 \begin{center}
 \includegraphics[width=\textwidth]{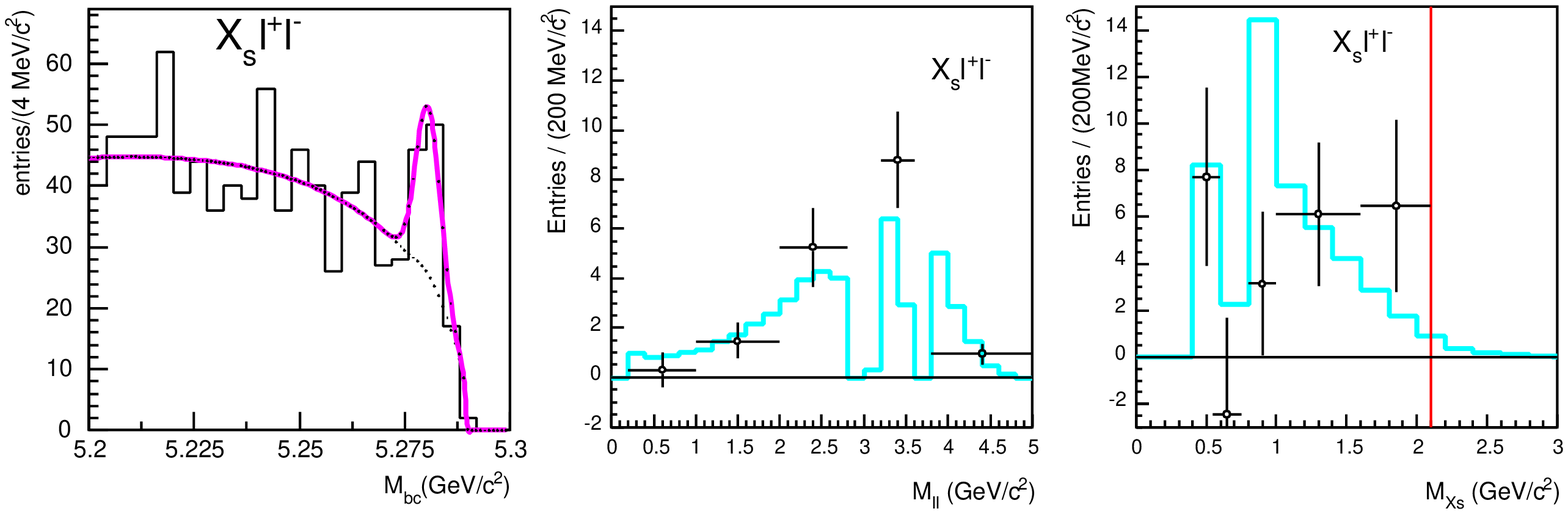}
 \caption{\it $\BtoXsll$ signal in $\Mbc$ (left) by Belle, and the
           $\Mll$ (middle) and $\MXs$ (right) distributions of the signal. }
 \label{fig:xsll}
 \end{center}
\end{figure}

The $42\fbinv$ dataset has been analyzed, and a signal of $48\pm11$
events is found in the $\Mbc$ distribution (Figure~\ref{fig:xsll}).
This corresponds to a $4.8\sigma$ significance and a branching fraction
of $(7.1\pm1.6^{+1.4}_{-1.2})\times10^{-6}$.  The result can be compared
with a SM prediction, $(4.2\pm0.7)\times10^{-6}$.  The distributions of
$\Mll$ and $\MXs$ are extracted from an analysis of the $\Mbc$
distributions for each bin of $\Mll$ and $\MXs$.  With the current
statistics, it is too early to compare with the SM predictions.  These
results will provide a stringent constraint on new physics.

%%%%%%%%%%%%%%%%%%%%%%%%%%%%%%%%%%%%%%%%%%%%%%%%%%%%%%%%%%%%%%%%%%%%%%
\section{Conclusion} %%%%%%%%%%%%%%%%%%%%%%%%%%%%%%%%%%%%%%%%%%%%%%%%%

After the successful start of the Belle and BaBar experiments,
$\btosgamma$ is now a mature topic; yet CLEO still provides the most
advanced results on the inclusive measurements although B-factories
which have a significantly larger datasets should catch up soon.  For
exclusive modes, results on both $\btosgamma$ and $\btodgamma$ by
B-factories already superseded CLEO.

The long awaited $\btosll$ measurements are finally available.  As
observed by two groups, the $\BtoKll$ signal is firmly established, and
$\BtoKstarll$ should be observed sooner or later.  The first inclusive
measurement of $\BtoXsll$ was performed by Belle, and will hopefully
become an important tool to find new physics in $B$ decays in the coming
years.

%%%%%%%%%%%%%%%%%%%%%%%%%%%%%%%%%%%%%%%%%%%%%%%%%%%%%%%%%%%%%%%%%%%%%%
\section{Acknowledgments} %%%%%%%%%%%%%%%%%%%%%%%%%%%%%%%%%%%%%%%%%%%%

I would like to express my thanks to the organizers of the Physics in
Collision conference.

%%%%%%%%%%%%%%%%%%%%%%%%%%%%%%%%%%%%%%%%%%%%%%%%%%%%%%%%%%%%%%%%%%%%%%
 %%%%%%%%%%%%%%%%%%%%%%%%%%%%%%%%%%%%%%%%%%%%%%%%

\end{document}